\documentclass[11pt, fleqn]{article}

\usepackage{xcolor, fullpage, amsmath, amsfonts, amsgen, amssymb, amsbsy, graphicx, appendix, physics, hyperref, bm, bbm}
\usepackage[numbers, sort&compress]{natbib}

\hypersetup{colorlinks = true, linkbordercolor = white, linkcolor = blue, citecolor = blue}
\newcommand{\equa}[1]{Eq.~(\ref{#1})} \newcommand{\equas}[1]{Eqs.~(\ref{#1})}
\newcommand{\equass}[2]{Eqs.~(\ref{#1})-(\ref{#2})}
\newcommand{\equasa}[2]{Eqs.~(\ref{#1}) and (\ref{#2})}

\newcommand{\eqn}[2]{\begin{gather}
#1
\label{#2}
\end{gather}
}

\title{\bf Gill's problem in a sandwiched porous slab}

\author{\normalsize {\bf Antonio Barletta$^a$\footnote{Correspondence: \tt antonio.barletta@unibo.it}~, Michele Celli$^a$, Stefano Lazzari$^b$, Pedro Vayssi\`ere Brand\~ao$^a$}\\[4pt]
{\small \it $^a$~Department of Industrial Engineering. Alma Mater Studiorum Universit\`a di Bologna.}\\ {\small \it Viale Risorgimento 2. 40136 Bologna. Italy.}\\[4pt]
{\small \it $^b$~Department of Architecture and Design. University of Genoa.}\\ 
{\small \it Stradone S. Agostino 37. 16123 Genoa. Italy.}}

\date{~}

\begin{document}

\maketitle

\begin{abstract}
The classical Gill's stability problem for the stationary and parallel buoyant flow in a vertical porous slab with impermeable and isothermal boundaries kept at different temperatures is reconsidered in a different perspective. A three-layer slab is studied instead of a homogeneous slab as in Gill's problem. The three layers have a symmetric configuration where the two external layers have a high thermal conductivity, while the core layer has a much lower conductivity. A simplified model is set up where the thermal conductivity ratio between the external layers and the internal core is assumed as infinite. It is shown that a flow instability in the sandwiched porous slab may arise with a sufficiently large Rayleigh number. It is also demonstrated that this instability coincides with that predicted in a previous analysis for a homogeneous porous layer with permeable boundaries, by considering the limiting case where the permeability of the external layers is much larger than that of the core layer. 
\end{abstract}



\section{Introduction}
The eventuality of convection heat transfer in a vertical porous layer saturated by a fluid is a topic widely studied over the last decades due to its potential interest for several applications. If the main practical use of this knowledge is for the thermal insulation of buildings and for devices such as breathing walls, the topic of convection in vertical porous layers may be important also for geophysics and for the design of filtration systems. The pioneering paper on this topic was published by \citet{gill1969proof}. This author provided a straightforward and rigorous proof that a vertical porous layer with a homogeneous structure and subjected to a side heating via isothermal boundaries cannot display a multicellular pattern of convection heat transfer whatever large is the Rayleigh number. This cornerstone result was obtained by modelling the flow through Darcy's law, the Boussinesq approximation and by assuming impermeable isothermal boundaries kept at different temperatures. The analysis carried out by \citet{gill1969proof} is grounded on the linear dynamics of perturbations superposed to the steady vertical buoyant flow in a conduction regime induced by the horizontal temperature gradient in the basic state. 

The paper by \citet{gill1969proof} is the starting point of several later studies including the nonlinear extension of the stability proof \citep{Stra88, Flavin1999}, as well as variants involving the Prandtl-Darcy flow model \citep{Rees1988} and the local thermal non-equilibrium within the saturated porous material \citep{Rees11, scott2013nonlinear}. 

The use of momentum balance models for the seepage flow in the porous layer that include the velocity Laplacian term shows up the possibility of a thermal instability and the emergence of convection patterns in the porous layer \citep{Chen20041257}. This feature is an expected consequence of a momentum balance model similar to that for a fluid clear of porous material. In fact, linear convective instability emerges when a vertical and infinitely tall slot of fluid is bounded by impermeable plane walls at different temperatures \citep{vest1969stability}. \citet{kwok1987stability} present experimental evidence that convective instability in a vertical porous layer may arise when a sufficiently large temperature difference is forced across the layer boundaries. These authors also propose possible explanations of the observed phenomenon based either on Brinkman's momentum transfer model or on the inclusion of variable viscosity effects \citep{kwok1987stability}.

The papers by \citet{Barletta2015} and by \citet{shankar2022gill} offer apparently different approaches showing up the emergence of convective instability in a vertical porous slab subjected to side heating. The first one \citep{Barletta2015} focusses on the same flow setup assumed by \citet{gill1969proof}, but turning the impermeability boundary conditions into conditions of permeability by imposing a hydrostatic pressure distribution at the boundaries. The second one \citep{shankar2022gill} considers a variant of Gill's setup where the boundaries are impermeable, but the porous medium is heterogeneous with a transverse continuous change of the permeability. Both \citet{Barletta2015} and \citet{shankar2022gill} proved that their variants of Gill's problem could lead to a condition of linear instability for sufficiently large Rayleigh numbers. 

The aim of this paper is to show that a common physical mechanism underlies the apparently different instabilities found by \citet{Barletta2015} and by \citet{shankar2022gill}. This task is achieved by envisaging a three-layer porous slab with impermeable isothermal boundaries kept at different temperatures. In this sandwiched porous slab, the two external layers are identical, while the core layer has different thermophysical properties. In particular, the external layers are considered as much more thermally conductive than the core. The basic buoyant flow in the internal core is identical to that devised by \citet{gill1969proof}. Although the basic state is stationary with a purely vertical velocity, the core layer has interfaces to the external layers which may allow for a horizontal flow contribution when the basic state is perturbed. This circumstance is a relaxation of Gill's impermeability condition at the boundaries, if just the core layer is considered. Such a setup is the link between the two studies by \citet{Barletta2015} and \citet{shankar2022gill}. In fact, the three-layer structure is effectively a horizontally heterogeneous medium with a piecewise constant permeability. Furthermore, the internal core is a homogeneous porous layer with permeability conditions at the bounding interfaces to the external layers.

In this study, it will be shown that the neutral stability condition is influenced by the permeability ratio between the external layers and the core layer and by the ratio between the thickness of the three-layer slab and that of the core layer. The role of such parameters in determining the stabilisation or destabilisation of the basic buoyant flow is identified by employing a numerical solution of the linear stability eigenvalue problem. This eigenvalue problem, formulated for the core layer, turns out to coincide with that solved by \citet{Barletta2015} in the asymptotic case where the external layers are much more permeable than the core. Hence, the neutral stability condition and the critical values of the wavenumber and of the Rayleigh number found by \citet{Barletta2015} are confirmed in the asymptotic case. Such an asymptotic condition turns out to be the most unstable relative to cases where the permeability ratio between the external layers and the core layer is finite. It will be shown that the stability eigenvalue problem studied by \citet{gill1969proof} is recovered in the opposite asymptotic condition where the permeability ratio between the external layers and the core layer tends to zero. This is an expected result as, in this asymptotic case, the interfaces between the core layer and the external layers are effectively impermeable, thus reproducing just the same Gill's boundary conditions for the core layer.

\section{Governing equations}
Let us consider a vertical porous slab with a sandwiched structure. The $x$ axis is horizontal and perpendicular to the slab, the $y$ axis is also horizontal and the $z$ axis is vertical, so that the gravitational acceleration is $\vb{g} = - g \, \vu{e}_z$, where $g$ is the modulus of $\vb{g}$ and $\vu{e}_z$ is the unit vector of the $z$ axis. 

\begin{figure}
\centering
\includegraphics[width=0.7\textwidth]{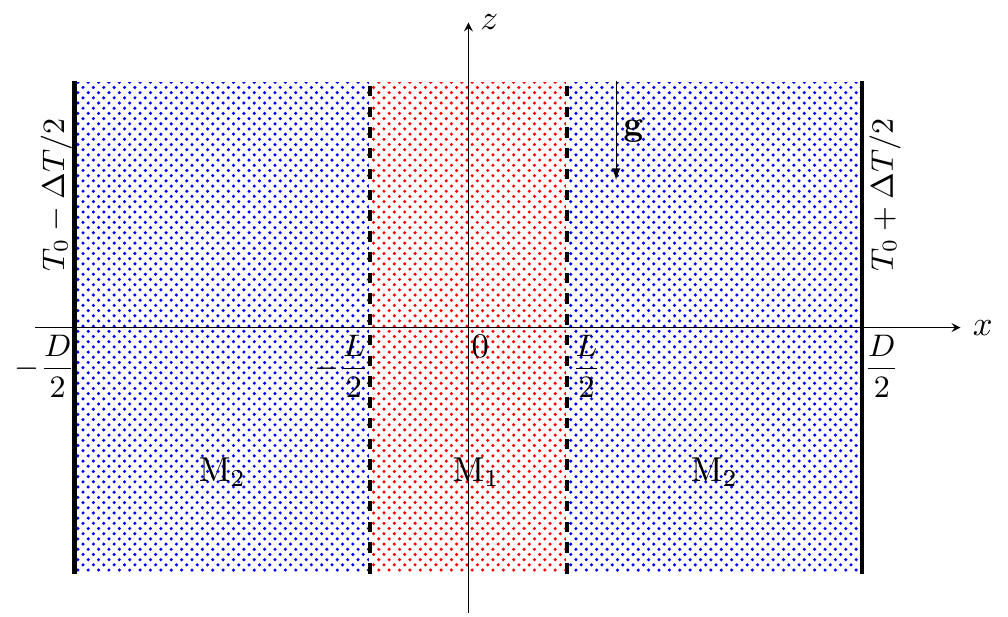}
\caption{\label{fig1}Two-dimensional cross-section of the sandwiched porous slab in the $xz$ plane. The $y$ axis is perpendicular to the plane of the figure.}
\end{figure}

As shown in Fig.~\ref{fig1},  the core region $-L/2 \le x \le L/2$ is a layer of a porous material ${\rm M}_1$, while the regions $-D/2 \le x < -L/2$ and $L/2 < x \le D/2$ are layers of a different porous material ${\rm M}_2$. For the sake of simplicity, we assume that the slab extends without bounds over the $y$ and $z$ directions. 
The external boundaries $x=\pm D/2$ are impermeable and isothermal at temperatures $T_0 \pm \Delta T/2$, where $T_0$ and $\Delta T$ are constants, with $\Delta T > 0$.

The flow model is based on the local mass, momentum and energy balance equations according to the Boussinesq approximation and to Darcy's law,
\eqn{
\div{\vb{u}_m} = 0,
\nonumber\\
\frac{\mu}{K_m}\, \vb{u}_m = - \grad{P_m} + \rho_0 g \beta \, \qty( T_m - T_0 )\, \vu{e}_z ,
\nonumber\\
\sigma_m\, \pdv{T_m}{t} + \vb{u}_m \vdot \grad{T_m} = \alpha_m \, \laplacian{T_m},\qquad m=1,2 ,
}{1}
where the index $m=1,2$ identifies the physical properties and the fields either in the core ${\rm M}_1$-layer or in the external ${\rm M}_2$-layers. Here, $t$ is time, $\vb{u}_m$ is the velocity field, $P_m$ is the dynamic pressure field, $T_m$ is the temperature field. The fluid saturating the three-layer slab is Newtonian with a dynamic viscosity $\mu$, a reference density $\rho_0$ and a coefficient of thermal expansion $\beta$. 

The average properties of the saturated porous media are, with $m=1,2$: the permeability $K_m$, the thermal diffusivity $\alpha_m$ and the heat capacity ratio $\sigma_m$. The latter ratio is evaluated by dividing the volumetric heat capacity of each saturated porous medium by the volumetric heat capacity of the fluid. We also recall that the average thermal diffusivities, $\alpha_1$ and $\alpha_2$, are defined through the ratio of the average thermal conductivity of the fluid saturated porous medium and the volumetric heat capacity of the fluid. The volumetric heat capacity of the fluid is the same for both media ${\rm M}_1$ and ${\rm M}_2$. Hence, the ratio $\alpha_2/\alpha_1$ is both a thermal diffusivity ratio and a thermal conductivity ratio.

\subsection{Pressure-temperature formulation}
Equation~(\ref{1}) can be expressed in a pressure-temperature formulation,
\eqn{
\laplacian{P_m} = \rho_0 g \beta \, \pdv{T_m}{z} ,
\nonumber\\
\sigma_m\, \pdv{T_m}{t} - \frac{K_m}{\mu}\, \grad{P_m} \vdot \grad{T_m} + \frac{\rho_0 g \beta K_m}{\mu} \, \qty( T_m - T_0 ) \, \pdv{T_m}{z} = \alpha_m \, \laplacian{T_m},\qquad m=1,2 .
}{2}

\subsection{Boundary and interface conditions}
The boundary and interface conditions are to be set at $x=\pm D/2$ and at $x=\pm L/2$, respectively,
\eqn{
x = \pm D/2 :  \qquad \pdv{P_2}{x} = 0 , \quad T_2 = T_0 \pm \frac{\Delta T}{2},
\nonumber \\
x = \pm L/2 :  \qquad P_1 = P_2 , \quad K_1\, \pdv{P_1}{x} = K_2\, \pdv{P_2}{x}, \quad T_1 = T_2, 
\quad \alpha_1\, \pdv{T_1}{x} = \alpha_2\, \pdv{T_2}{x} .
}{3}
In particular, the interface conditions reflect the momentum and energy balance across the planes at $x = \pm L/2$. They express the continuity of the pressure, of the normal component of the velocity, of the temperature and of the normal component of the heat flux density. The last interface condition (\ref{3}) should employ the thermal conductivities instead of the thermal diffusivities, but we already pointed out that the ratio $\alpha_2/\alpha_1$ coincides with the thermal conductivity ratio.

\subsection{Dimensionless formulation}
The mathematical model can be reformulated in dimensionless terms by scaling time, coordinates and fields as,
\eqn{
t^* = \frac{t}{\sigma_1 L^2/\alpha_1},\quad \qty(x^*, y^*, z^*) = \frac{(x,y,z)}{L}, 
\quad 
\vb{u}^*_m = \frac{\vb{u}_m}{\alpha_1/L}, 
\nonumber\\ 
P^*_m = \frac{P_m}{\mu \alpha_1/K_1}, \quad T^*_m = \frac{T_m - T_0}{\Delta T},
}{4}
with the dimensionless parameters,
\eqn{
R = \frac{\rho_0 g \beta \Delta T K_1 L}{\mu \alpha_1}, \quad a = \frac{D}{L}, \quad \xi =\frac{K_2}{K_1} , \quad \gamma = \frac{\alpha_2}{\alpha_1} \quad \tau = \frac{\sigma_2}{\sigma_1} .
}{5}
Here, $a > 1$ and $R$ is the Darcy-Rayleigh number, hereafter called Rayleigh number for conciseness. Equations~(\ref{4}) and (\ref{5}) allow one to rewrite \equa{2} in a dimensionless form for the medium ${\rm M}_1$,
\eqn{
\laplacian{P_1} = R \, \pdv{T_1}{z} ,
\nonumber\\
\pdv{T_1}{t} - \grad{P_1} \vdot \grad{T_1} + R\,T_1 \, \pdv{T_1}{z} = \laplacian{T_1},
}{6}
and for the medium ${\rm M}_2$,
\eqn{
\laplacian{P_2} = R \, \pdv{T_2}{z} ,
\nonumber\\
\tau\, \pdv{T_2}{t} - \xi\, \grad{P_2} \vdot \grad{T_2} + R \xi\,T_2 \, \pdv{T_2}{z} = \gamma\, \laplacian{T_2},
}{7}
while \equa{3} reads
\eqn{
x = \pm a/2 :  \qquad \pdv{P_2}{x} = 0 , \quad T_2 = \pm \frac{1}{2},
\nonumber \\
x = \pm 1/2 :  \qquad P_1 = P_2 , \quad \pdv{P_1}{x} = \xi\, \pdv{P_2}{x}, \quad T_1 = T_2, 
\quad \pdv{T_1}{x} = \gamma\, \pdv{T_2}{x} .
}{8}
In \equass{6}{8} and in the forthcoming analysis, the dimensionless fields, coordinates and time are denoted without the asterisks for simplicity of notation. This will not cause any ambiguity as we will only deal with dimensionless expressions, except when explicitly declared.

It is also worth saying that the second \equa{1} is rewritten in a dimensionless form as
\eqn{
\vb{u}_1 = - \grad{P_1} + R\, T_1\, \vu{e}_z , 
\nonumber \\
\vb{u}_2 = - \xi\, \grad{P_2} + R \xi\, T_2\, \vu{e}_z , 
}{9}
for the media ${\rm M}_1$ and ${\rm M}_2$, respectively. 

\section{The basic stationary flow}
A basic stationary flow solution of \equass{6}{8} can be found such that
\eqn{
\bar{P}_1 = 0 = \bar{P}_2, \quad \pdv{\bar{T}_1}{y} = 0 = \pdv{\bar{T}_2}{y}, 
}{10}
where the bar over the fields stands for basic state.
Equation~(\ref{10}) describes a two-dimensional, $y$ independent, flow regime where the pressure locally coincides with the hydrostatic pressure or, equivalently, the dynamic pressure is everywhere zero. As a consequence of \equa{10}, \equass{6}{8} allow one to infer that, in the three-layer slab, the temperature profile $\bar{T}$ is a piecewise linear function of $x$ independent of $y$ and $z$. In fact, in the basic state, we have
\eqn{
\bar{T}_1 = \dfrac{\gamma}{\gamma + a - 1}\, x, \qquad 
\bar{T}_2 = 
\begin{cases}
\dfrac{1}{\gamma + a -1}\, x - \dfrac{\gamma - 1}{2 \qty(\gamma + a - 1)}, \quad -\dfrac{a}{2} \le x < - \dfrac{1}{2},\\[12pt]
\dfrac{1}{\gamma + a -1}\, x + \dfrac{\gamma - 1}{2 \qty(\gamma + a - 1)}, \quad \dfrac{1}{2} < x \le \dfrac{a}{2}.
\end{cases}
}{11}
Another consequence of \equa{10}, is that \equa{9} yields
\eqn{
\bar{\vb{u}}_1 = \qty(0, 0, R\, \bar{T}_1), \quad \bar{\vb{u}}_2 = \qty(0, 0, R\xi\, \bar{T}_2),
}{12}
which describes a purely vertical flow driven only by the buoyancy force. If $\bar{T}$ varies continuously along the range $-a/2 \le x \le a/2$, this is not the case for the vertical velocity component. Such a feature is a consequence of the different permeabilities of the porous media ${\rm M}_1$ and ${\rm M}_2$, so that continuity of the velocity profile occurs only for the special case $\xi=1$.

\begin{figure}
\centering
\includegraphics[width=0.9\textwidth]{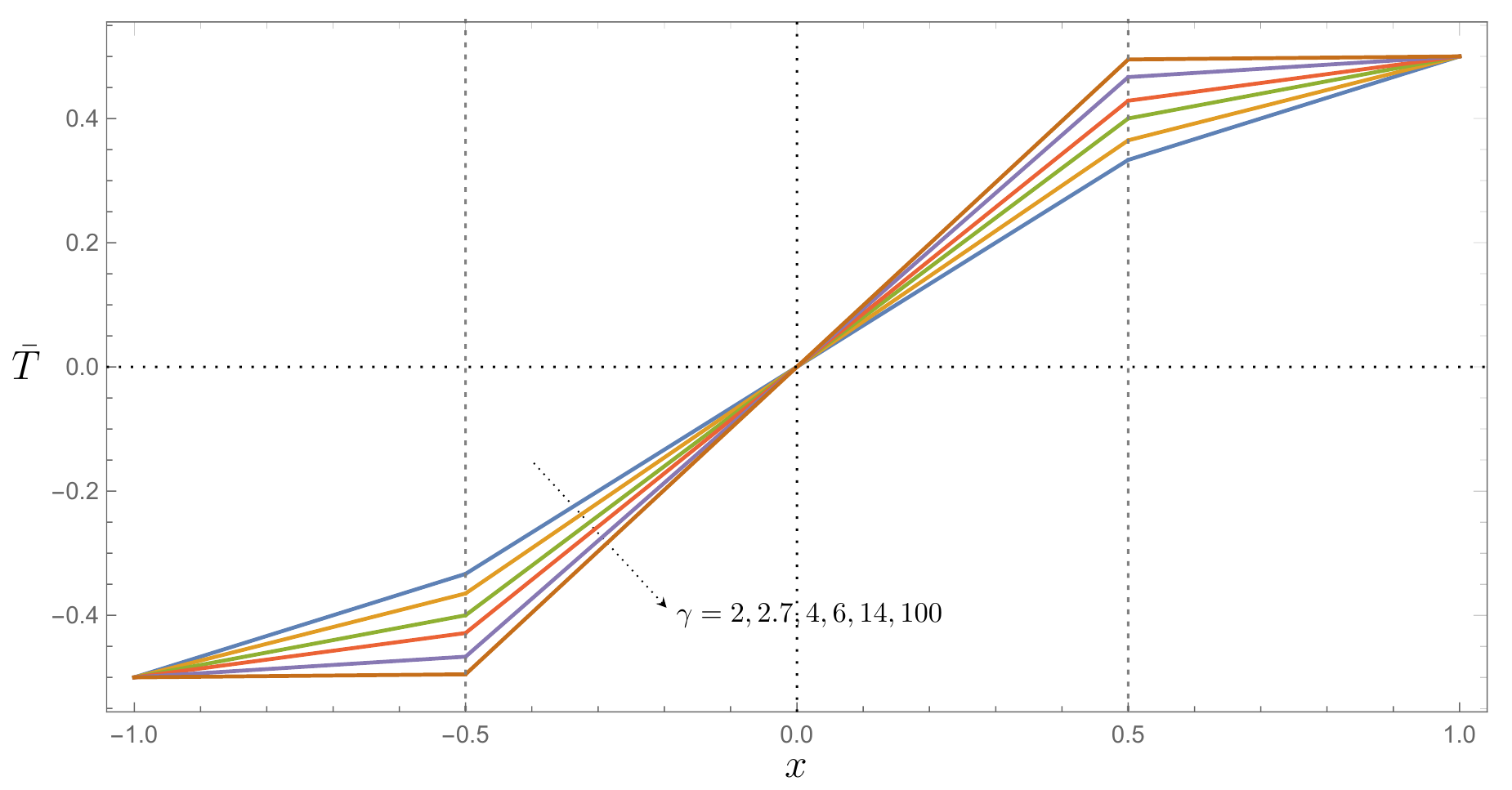}
\caption{\label{fig2}Plots of $\bar{T}$ versus $x$ for $a=2$ and different values of $\gamma$.}
\end{figure}

Figure~\ref{fig2} illustrates the basic solution by showing some plots of $\bar{T}$ versus $x$ over the whole range $-a/2 \le x \le a/2$. The geometrical ratio $a=2$ is prescribed, while different values of the conductivity ratio $\gamma > 1$ are considered. A focus on the case $\gamma > 1$ has been made since, as it will be discussed in Section~\ref{infcon}, our aim is to explore a condition of extremely high values of $\gamma$. It can be stressed that the case $\gamma = 100$ yields a temperature distribution which is almost uniform in the external layers ($-a/2 \le x < -1/2$ and $1/2 < x \le a/2$). This behaviour is characteristic of the asymptotic case $\gamma \to \infty$.


\subsection{Infinite thermal conductivity ratio}\label{infcon}

There are several practical cases where the limit $\gamma \to \infty$ is a fairly appropriate condition. Porous media with a very high thermal conductivity are the metal foams, often employed in the design of heat exchangers, while the low conductivity inner core can be devised as a slab of any thermal insulation material employed in the building industry. In the forthcoming stability analysis of the basic buoyant flow, we will focus on this asymptotic condition. It is worth emphasising that the limit $\gamma \to \infty$ of \equa{11} yields
\eqn{
\bar{T}_1 = x, \qquad 
\bar{T}_2 = 
\begin{cases}
- \dfrac{1}{2}, \quad -\dfrac{a}{2} \le x < - \dfrac{1}{2},\\[12pt]
\dfrac{1}{2}, \quad \dfrac{1}{2} < x \le \dfrac{a}{2}.
\end{cases}
}{14}
The basic state (\ref{14}) for the core layer ${\rm M}_1$ is exactly that considered in the papers by \citet{gill1969proof} and by \citet{Barletta2015}. Finally, we reckon that also the governing equations, the boundary and the interface conditions (\ref{7}) and (\ref{8}) undergo a drastic simplification,
\eqn{
\laplacian{P_2} = R \, \pdv{T_2}{z} ,
\quad
\laplacian{T_2} = 0,
}{15}
with
\eqn{
x = \pm a/2 :  \qquad \pdv{P_2}{x} = 0 , \quad T_2 = \pm \frac{1}{2},
\nonumber \\
x = \pm 1/2 :  \qquad P_1 = P_2 , \quad \pdv{P_1}{x} = \xi\, \pdv{P_2}{x}, \quad T_1 = T_2, 
\quad \pdv{T_2}{x} = 0 .
}{16}

\section{Linearised perturbation dynamics}
The onset of instability is studied by perturbing the basic state,
\eqn{
P_m = \bar{P}_m + \epsilon \, \hat{P}_m , \quad T_m = \bar{T}_m + \epsilon \, \hat{T}_m , \quad m=1,2,
}{17}
where $\epsilon>0$ is a perturbation parameter and the hat identifies the perturbation contributions to the pressure and temperature fields. The linear stability analysis is carried out by substituting \equa{17} into \equas{6}, (\ref{15}) and (\ref{16}) and by neglecting the terms $\order{\epsilon^2}$. We obtain, for the inner layer ${\rm M}_1$,
\eqn{
\laplacian{\hat{P}_1} = R \, \pdv{\hat{T}_1}{z} 
\qc
\pdv{\hat{T}_1}{t} - \pdv{\hat{P}_1}{x} + R\,x \, \pdv{\hat{T}_1}{z} = \laplacian{\hat{T}_1},
}{18}
and, for the external layers ${\rm M}_2$,
\eqn{
\laplacian{\hat{P}_2} = R \, \pdv{\hat{T}_2}{z} 
\qc
\laplacian{\hat{T}_2} = 0,
}{19}
with
\eqn{
x = \pm a/2 :  \qquad \pdv{\hat{P}_2}{x} = 0 , \quad \hat{T}_2 = 0,
\nonumber \\
x = \pm 1/2 :  \qquad \hat{P}_1 = \hat{P}_2 , \quad \pdv{\hat{P}_1}{x} = \xi\, \pdv{\hat{P}_2}{x}, \quad \hat{T}_1 = \hat{T}_2, 
\quad \pdv{\hat{T}_2}{x} = 0 .
}{20}
Here, the features of the basic buoyant flow, defined by \equasa{10}{14}, have been employed.

\subsection{Normal mode analysis}
A normal mode expression of the perturbations $(\hat{P}_m, \hat{T}_m)$ is given by
\eqn{
\left[
\begin{array}{c}
\hat{P}_m \\ 
\hat{T}_m
\end{array}
\right]
=
\left[
\begin{array}{c}
{f}_m(x) \\ 
{h}_m(x)
\end{array}
\right]
e^{i \qty(k_y y + k_z z - \omega t)}, \quad \text{with\ } k_y \in \mathbb{R}, k_z \in \mathbb{R}, \omega \in \mathbb{C}, \quad m = 1, 2,
}{21}
where $\vb{k} = \qty(0, k_y, k_z)$ is the wave vector and $\omega$ is a complex angular frequency. The wave number is a positive quantity defined as the modulus of the wave vector, $k = |\vb{k}|$. The angular frequency is the real part of $\omega$, while the temporal growth rate is the imaginary part of $\omega$. By substituting \equa{21} into \equass{18}{20}, one obtains for ${\rm M}_1$
\eqn{
f''_1 - k^2\, f_1 - i k_z R \, h_1 = 0 ,
\nonumber\\
h''_1 - \qty(k^2 - i \omega + i k_z R x)\, h_1 + f'_1 = 0,
}{22}
and for the external ${\rm M}_2$ layers
\eqn{
f''_2 - k^2\, f_2 - i k_z R \, h_2 = 0 ,
\nonumber\\
h''_2 - k^2\, h_2 = 0,
}{23}
with the boundary and interface conditions
\eqn{
x = \pm a/2 :  \qquad f'_2 = 0 , \quad h_2 = 0,
\nonumber \\
x = \pm 1/2 :  \qquad f_1 = f_2 , \quad f'_1 = \xi\, f'_2, \quad h_1 = h_2, 
\quad h'_2 = 0 .
}{24}
We can get rid of the dependence on the orientation of the wave vector $\vb{k}$ in \equass{22}{24} by defining a rescaled Rayleigh number, $S$, such that
\eqn{
k S = k_z R.
}{25}
Hence, \equasa{22}{23} read
\eqn{
f''_1 - k^2\, f_1 - i k S \, h_1 = 0 ,
\nonumber\\
h''_1 - \qty(k^2 - i \omega + i k S x)\, h_1 + f'_1 = 0,
}{26}
and
\eqn{
f''_2 - k^2\, f_2 - i k S \, h_2 = 0 ,
\nonumber\\
h''_2 - k^2\, h_2 = 0,
}{27}
respectively.

\subsection{Eigenvalue problem for the core porous layer}
Function $h_2$ is a solution of the second \equa{27} which, on account of \equa{24}, must satisfy the boundary conditions $h_2(\pm a/2)=0$. Then, it can be expressed as
\eqn{
h_2 = 
\begin{cases}
C_h  \sinh\!\qty(k \, \dfrac{a + 2 x}{2}), \quad -\dfrac{a}{2} \le x < - \dfrac{1}{2}, \\[12pt]
\tilde{C}_h  \sinh\!\qty(k \, \dfrac{a - 2 x}{2}), \quad \dfrac{1}{2} < x \le \dfrac{a}{2},
\end{cases}
}{28}
where $\qty(C_h, \tilde{C}_h)$ are integration constants. However, \equa{24} prescribes also the interface conditions $h'_2(\pm 1/2) = 0$. On account of \equa{28}, such conditions can be satisfied for $k>0$ only with $C_h = 0$ and $\tilde{C}_h = 0$. Hence, one has
\eqn{
h_2 = 0 .
}{29}
Function $f_2$ must be a solution of the first \equa{27} with $h_2 = 0$. Furthermore, on account of \equa{24}, $f_2$ satisfies the boundary conditions $f'_2(\pm a/2)=0$, so that
\eqn{
f_2 = 
\begin{cases}
C_f  \cosh\!\qty(k \, \dfrac{a + 2 x}{2}), \quad -\dfrac{a}{2} \le x < - \dfrac{1}{2}, \\[12pt]
\tilde{C}_f  \cosh\!\qty(k \, \dfrac{a - 2 x}{2}), \quad \dfrac{1}{2} < x \le \dfrac{a}{2},
\end{cases}
}{30}
where $\qty(C_f, \tilde{C}_f)$ are integration constants.\\ Thus, from \equa{30}, the interface conditions $f_1(- 1/2) = f_2(- 1/2)$ and $f'_1(- 1/2) = \xi\, f'_2(- 1/2)$ given by \equa{24} yield
\eqn{
f_1\qty(-\dfrac{1}{2}) = C_f  \cosh\!\qty(k \, \dfrac{a - 1}{2}), \quad f'_1\qty(-\dfrac{1}{2}) = C_f \xi k \sinh\!\qty(k \, \dfrac{a - 1}{2}) .
}{31}
By eliminating $C_f$, \equa{31} yields
\eqn{
f'_1\qty(-\frac{1}{2}) - \xi k \tanh\!\qty(k \, \dfrac{a - 1}{2})\, f_1\qty(-\frac{1}{2}) = 0 .
}{32}
By employing the same method, the interface conditions $f_1(1/2) = f_2(1/2)$ and $f'_1(1/2) = \xi\, f'_2(1/2)$ given by \equa{24} yield
\eqn{
f_1\qty(\frac{1}{2}) = \tilde{C}_f  \cosh\!\qty(k \, \dfrac{a - 1}{2}), \quad f'_1\qty(\frac{1}{2}) = - \tilde{C}_f \xi k \sinh\!\qty(k \, \dfrac{a - 1}{2}) .
}{33}
By eliminating $\tilde{C}_f$, \equa{33} yields
\eqn{
f'_1\qty(\dfrac{1}{2}) + \xi k \tanh\!\qty(k \, \dfrac{a - 1}{2})\, f_1\qty(\dfrac{1}{2}) = 0 .
}{34}
On account of  \equas{26}, (\ref{32}) and (\ref{34}), one can conclude that the linear stability analysis can be now formulated by focussing just on the inner core region $-1/2 \le x \le 1/2$ and solving the eigenvalue problem
\eqn{
f''_1 - k^2\, f_1 - i k S \, h_1 = 0 ,
\nonumber\\
h''_1 - \qty(k^2 - i \omega + i k S x)\, h_1 + f'_1 = 0,
\nonumber\\
x = \pm \dfrac{1}{2}: \quad f'_1 \pm \xi k \tanh\!\qty(k \, \dfrac{a - 1}{2})\, f_1 = 0,  \quad h_1 = 0. 
}{35}
After having solved \equa{35} and determined $f_1$, one can determine also $f_2$ as
\eqn{
f_2 = 
\begin{cases}
f_1\qty(-\dfrac{1}{2}) \qty[\cosh\!\qty(k \, \dfrac{a - 1}{2})]^{-1}  \cosh\!\qty(k \, \dfrac{a + 2 x}{2}), \quad -\dfrac{a}{2} \le x < - \dfrac{1}{2}, \\[12pt]
f_1\qty(\dfrac{1}{2}) \qty[\cosh\!\qty(k \, \dfrac{a - 1}{2})]^{-1}  \cosh\!\qty(k \, \dfrac{a - 2 x}{2}), \quad \dfrac{1}{2} < x \le \dfrac{a}{2},
\end{cases}
}{36}
where \equas{30}, (\ref{31}) and (\ref{33}) have been used.

\subsection{Features of the stability eigenvalue problem}
We note that \equasa{32}{34} are boundary conditions of the third kind for $f_1$. However, they are of a special type as the coefficient of the $f_1$ term depends on the wave number $k$. A similar circumstance occurs in the linear stability eigenvalue problems solved by \citet{rees2013effect} and by \citet{mohammad2017effect}, where third-kind boundary conditions with a $k$-dependent coefficient were found for the temperature disturbances. In particular, in \citet{rees2013effect}, Prats' problem in a horizontal porous layer \citep{Prats} was reconsidered by studying the effect of the nonzero thermal resistance of the horizontal boundary walls. Thus, from a mathematical viewpoint, the circumstances devised by these authors are comparable with those considered here and leading to \equa{35}. In fact, we are studying a finite and nonzero hydraulic (instead of thermal) resistance of the external porous layers. We mention that the case of third-kind boundary conditions for the temperature were considered in the analysis of Gill's problem for a vertical plane slab with permeable boundaries by \citet{barletta2017unstable}, while third-kind boundary conditions for the pressure were investigated by \citet{barletta2020stability}. However, in \citet{barletta2017unstable, barletta2020stability}, the prescribed third-kind conditions feature constant $k$-independent coefficients, unlike the case defined by \equa{35}. We mention that third-kind boundary conditions with constant coefficients were also predicted for the pressure field by \citet{nygaard2010onset}.

\section{Discussion of the results}
The basis for developing the stability analysis is the eigenvalue problem (\ref{35}). Some important characteristics of the onset of instability can be gathered by exploring three significant asymptotic cases.

\subsection{The limit $\xi \to \infty$}\label{csinf}
The limiting condition $\xi \to \infty$ embodies the case where the external ${\rm M}_2$ layers are much more permeable than the core ${\rm M}_1$ layer. We note that the example of metal foams for the ${\rm M}_2$ layers is quite close to this condition as the metal foams are generally endowed with a large permeability. In any case, we assume that the large permeability of the ${\rm M}_2$ layers is not so large as to suppress the validity of Darcy's law. In this case, the only change in the stability eigenvalue problem is a drastic simplification of the boundary conditions. Equation~(\ref{35}) thus reads 
\eqn{
f''_1 - k^2\, f_1 - i k S \, h_1 = 0 ,
\nonumber\\
h''_1 - \qty(k^2 - i \omega + i k S x)\, h_1 + f'_1 = 0,
\nonumber\\
x = \pm \dfrac{1}{2}: \quad f_1 = 0,  \quad h_1 = 0. 
}{37}
We emphasise that \equa{37} coincides with the stability eigenvalue problem for a homogeneous vertical porous layer with permeable boundaries solved by \citet{Barletta2015}. This is an important result as the transition to instability in the limiting case $\xi \to \infty$ is defined by the neutral stability data obtained and discussed by \citet{Barletta2015}. On taking the limit $\xi \to \infty$, one loses also the dependence on $a$.

\subsection{The limits $\xi \to 0$ or $a \to 1$}\label{csizero}
An asymptotic case completely different from that discussed in Section~\ref{csinf} is defined by the limit $\xi \to 0$. This limit describes a situation where the ${\rm M}_2$ layers are much less permeable than the core ${\rm M}_1$ layer. Strictly speaking, the external layers are quite close to impermeability if compared with the core layer. In this case, \equa{35} drastically simplifies to
\eqn{
f''_1 - k^2\, f_1 - i k S \, h_1 = 0 ,
\nonumber\\
h''_1 - \qty(k^2 - i \omega + i k S x)\, h_1 + f'_1 = 0,
\nonumber\\
x = \pm \dfrac{1}{2}: \quad f'_1 = 0,  \quad h_1 = 0. 
}{38}
The stability eigenvalue problem (\ref{38}) coincides with that analysed by \citet{gill1969proof}. This feature is completely unsurprising as \citet{gill1969proof} investigated the possible onset of instability in a homogeneous porous layer with impermeable isothermal boundaries kept at different temperatures. \citet{gill1969proof} proved that no instability is possible in this case.

It is significant that \equa{38} represents also the limiting case $a \to 1$ for any finite $\xi$. This result is expected as $a \to 1$ means that the thickness of the ${\rm M}_2$ layers tends to $0$. Thus, one has again a homogeneous porous layer, namely the ${\rm M}_1$ layer, with the impermeable boundaries devised in \citet{gill1969proof}.

\subsection{The limit $a \to \infty$}\label{ainf}
When the external ${\rm M}_2$ layers have an extremely large thickness, so that $D \gg L$, we have in mind the core ${\rm M}_1$ layer surrounded by infinite ${\rm M}_2$ media. This condition yields the limit $a \to \infty$ and the stability eigenvalue problem (\ref{35}) simplifies to
\eqn{
f''_1 - k^2\, f_1 - i k S \, h_1 = 0 ,
\nonumber\\
h''_1 - \qty(k^2 - i \omega + i k S x)\, h_1 + f'_1 = 0,
\nonumber\\
x = \pm \dfrac{1}{2}: \quad f'_1 \pm \xi k\, f_1 = 0,  \quad h_1 = 0. 
}{39}

\begin{figure}
\centering
\includegraphics[width=0.5\textwidth]{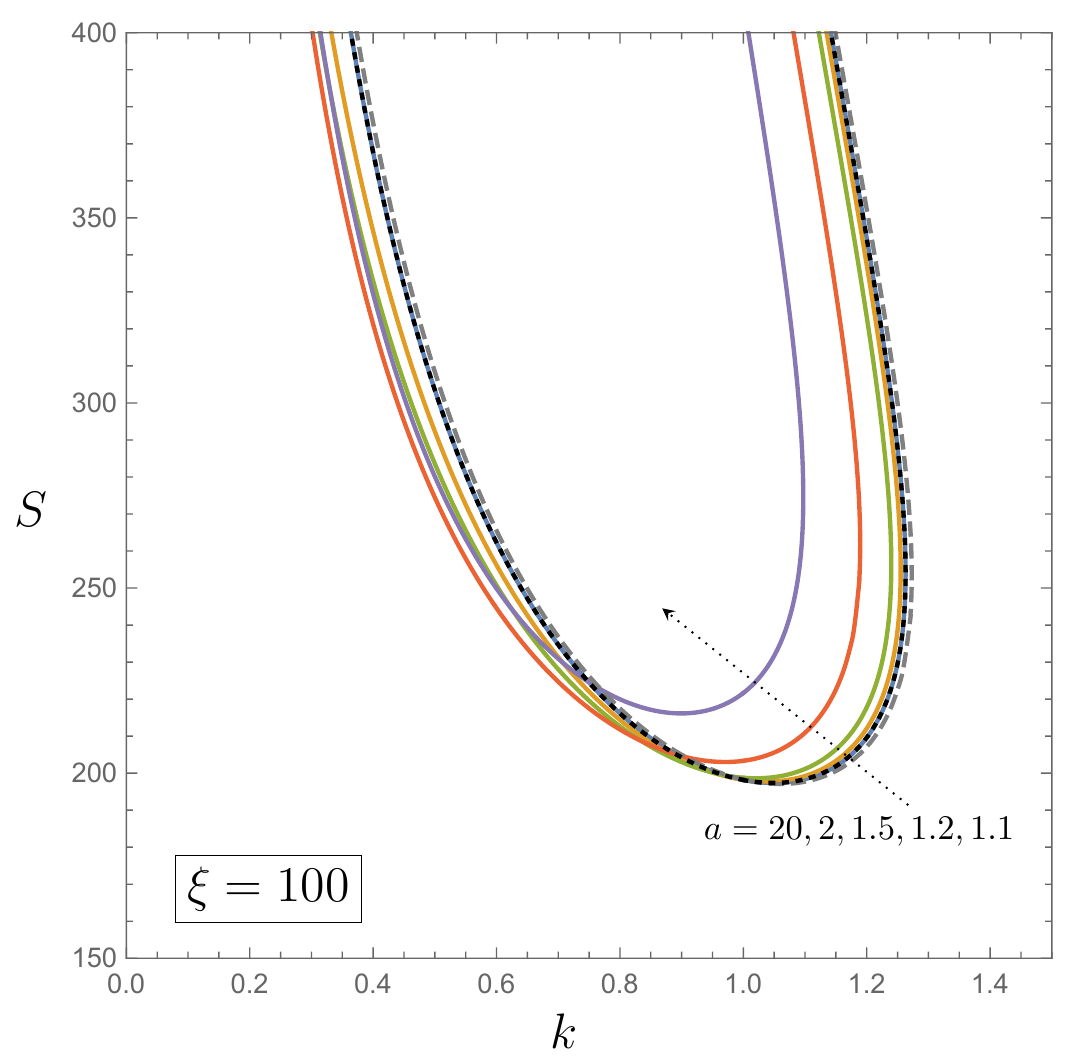}\includegraphics[width=0.5\textwidth]{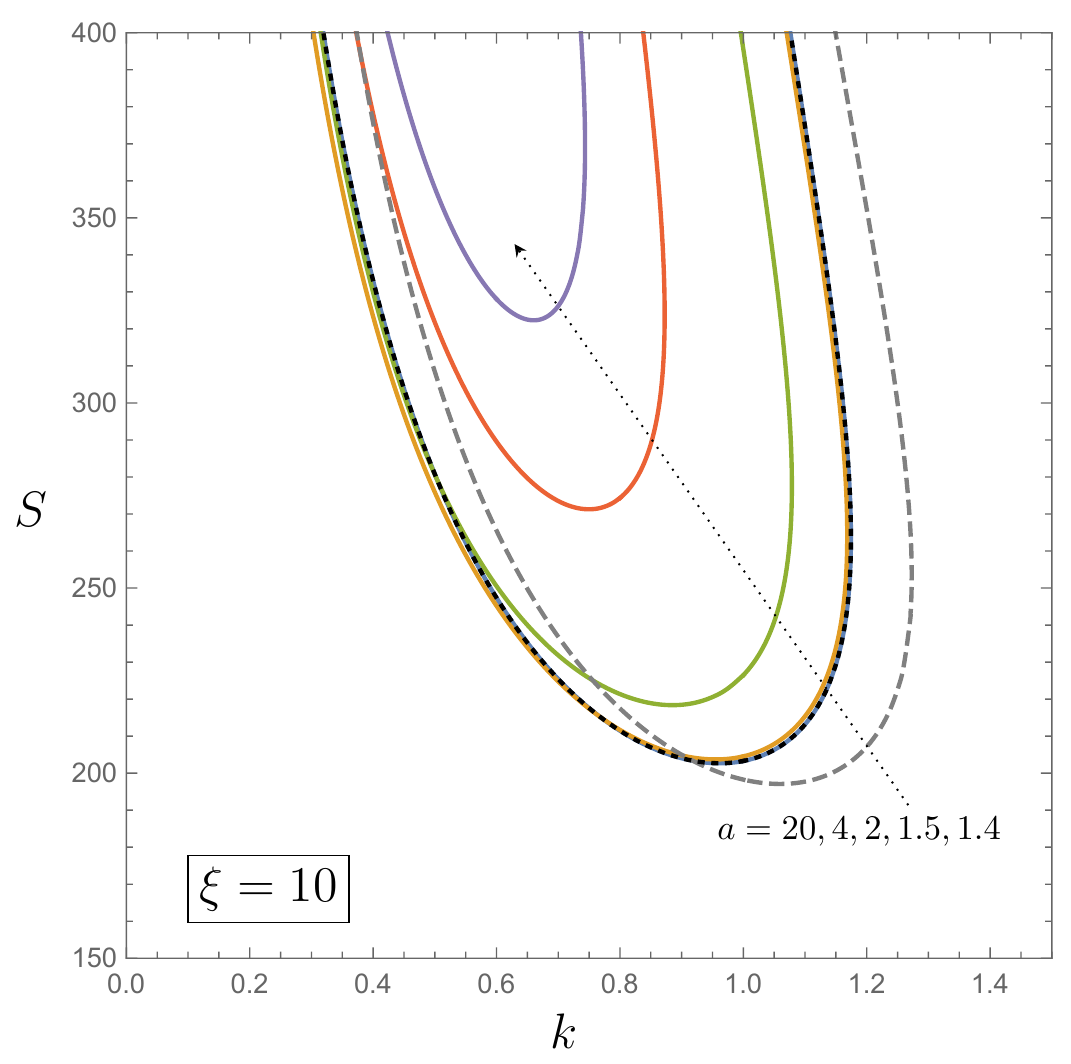}
\caption{\label{fig3}Neutral stability curves in the $(k,S)$ plane for $\xi=100$ and $\xi = 10$ with different values of $a$ (solid lines). The dashed grey line corresponds to the limit $\xi \to \infty$, while the dotted black lines describe the limit $a \to \infty$.}
\end{figure}

\begin{figure}
\centering
\includegraphics[width=0.5\textwidth]{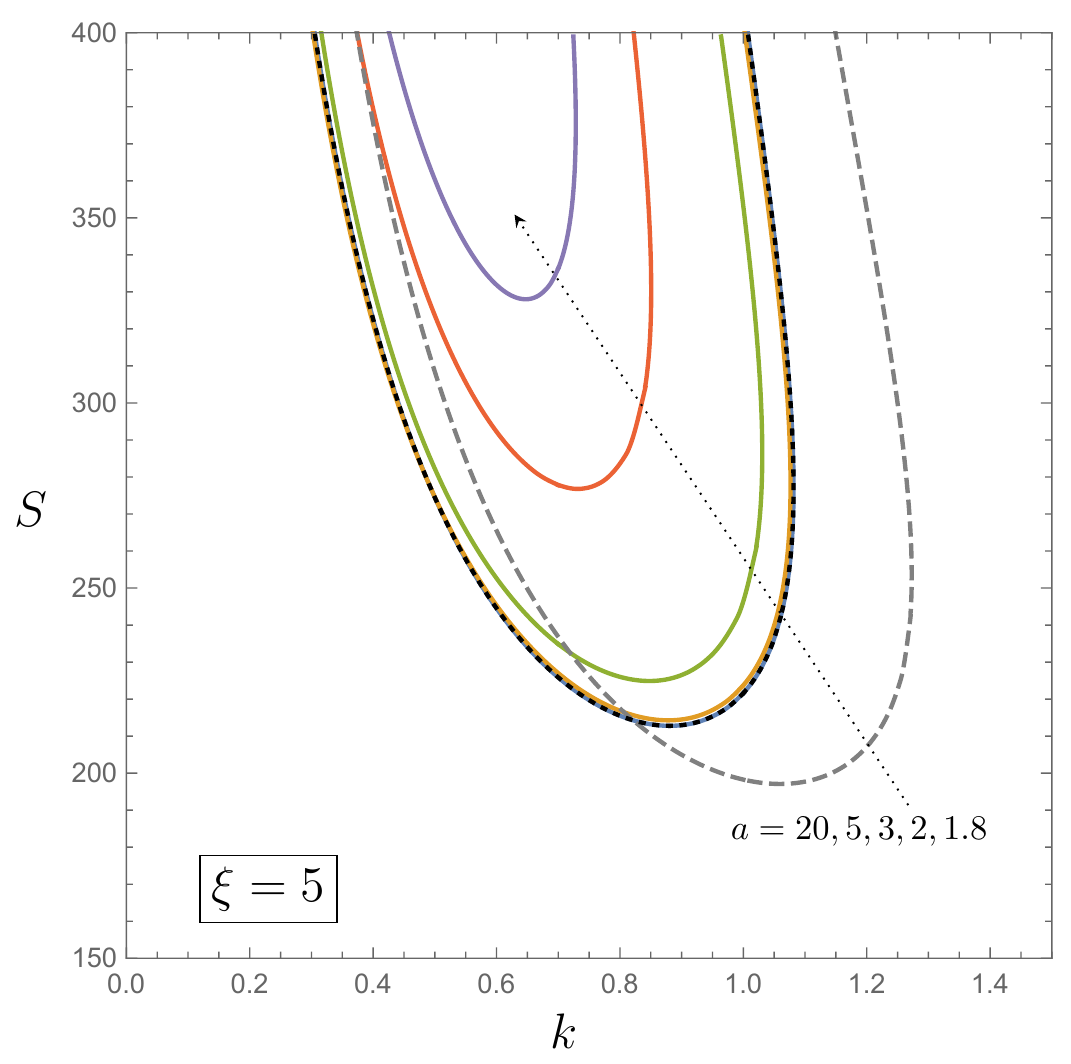}\includegraphics[width=0.5\textwidth]{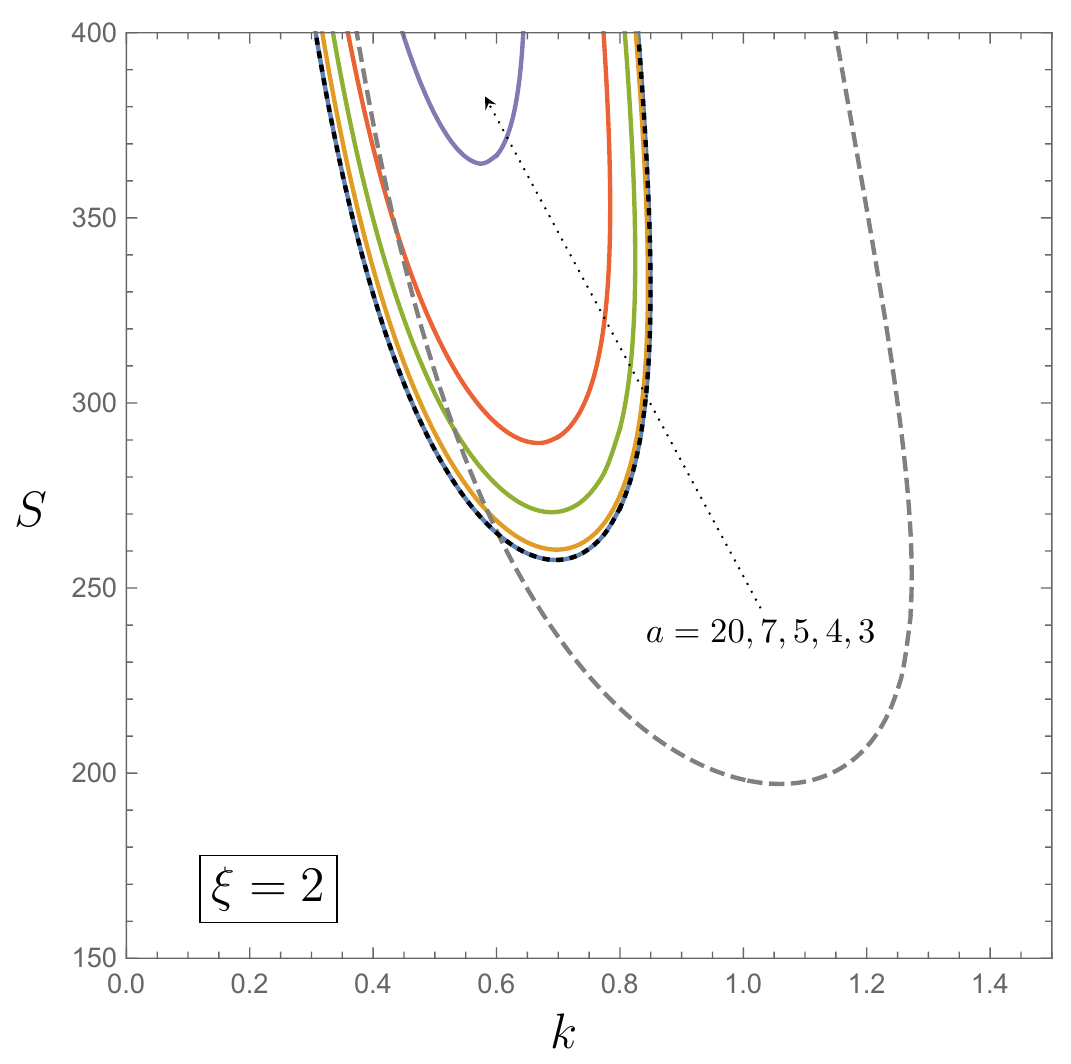}
\caption{\label{fig4}Neutral stability curves in the $(k,S)$ plane for $\xi=5$ and $\xi = 2$ with different values of $a$ (solid lines). The dashed grey line corresponds to the limit $\xi \to \infty$, while the dotted black lines describe the limit $a \to \infty$.}
\end{figure}

\begin{figure}
\centering
\includegraphics[width=0.5\textwidth]{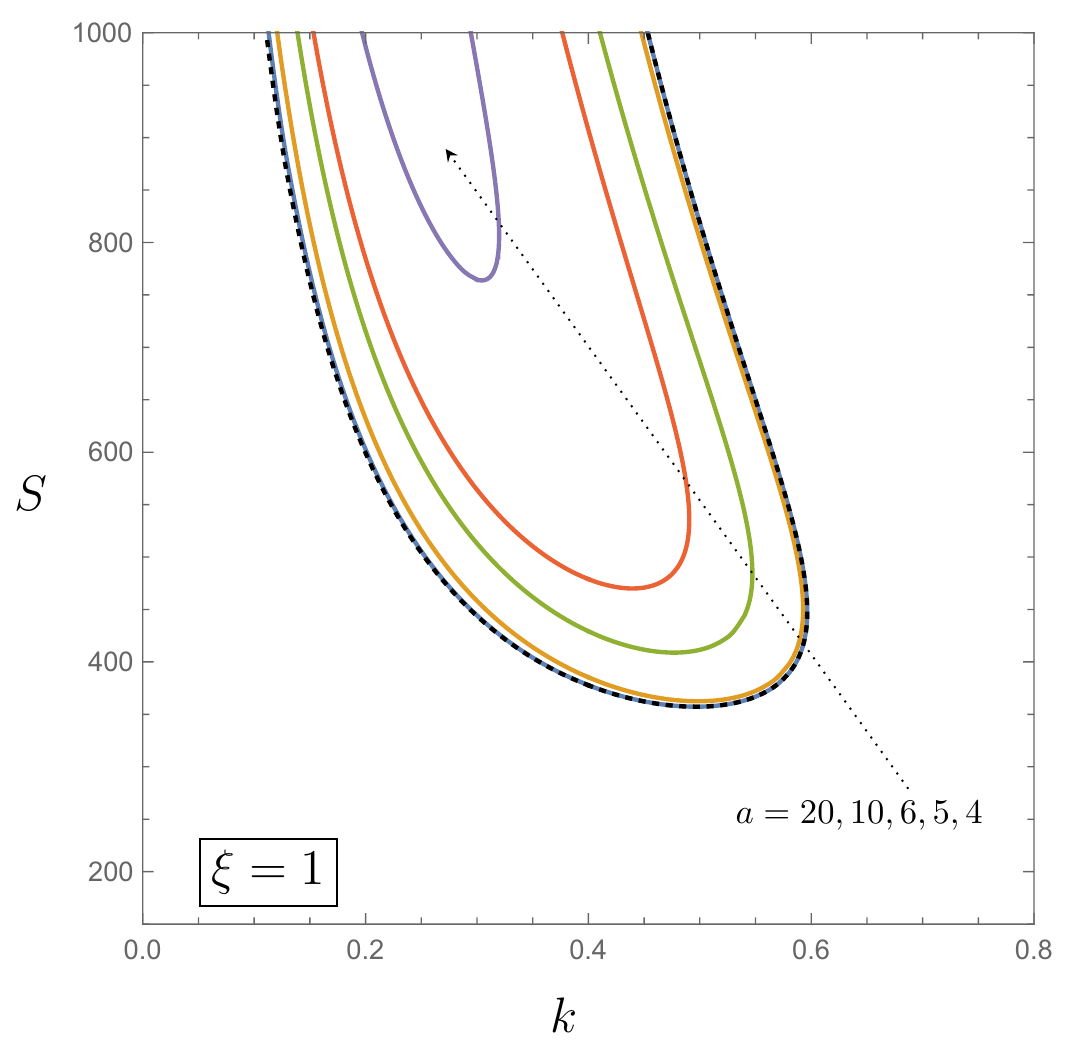}\includegraphics[width=0.5\textwidth]{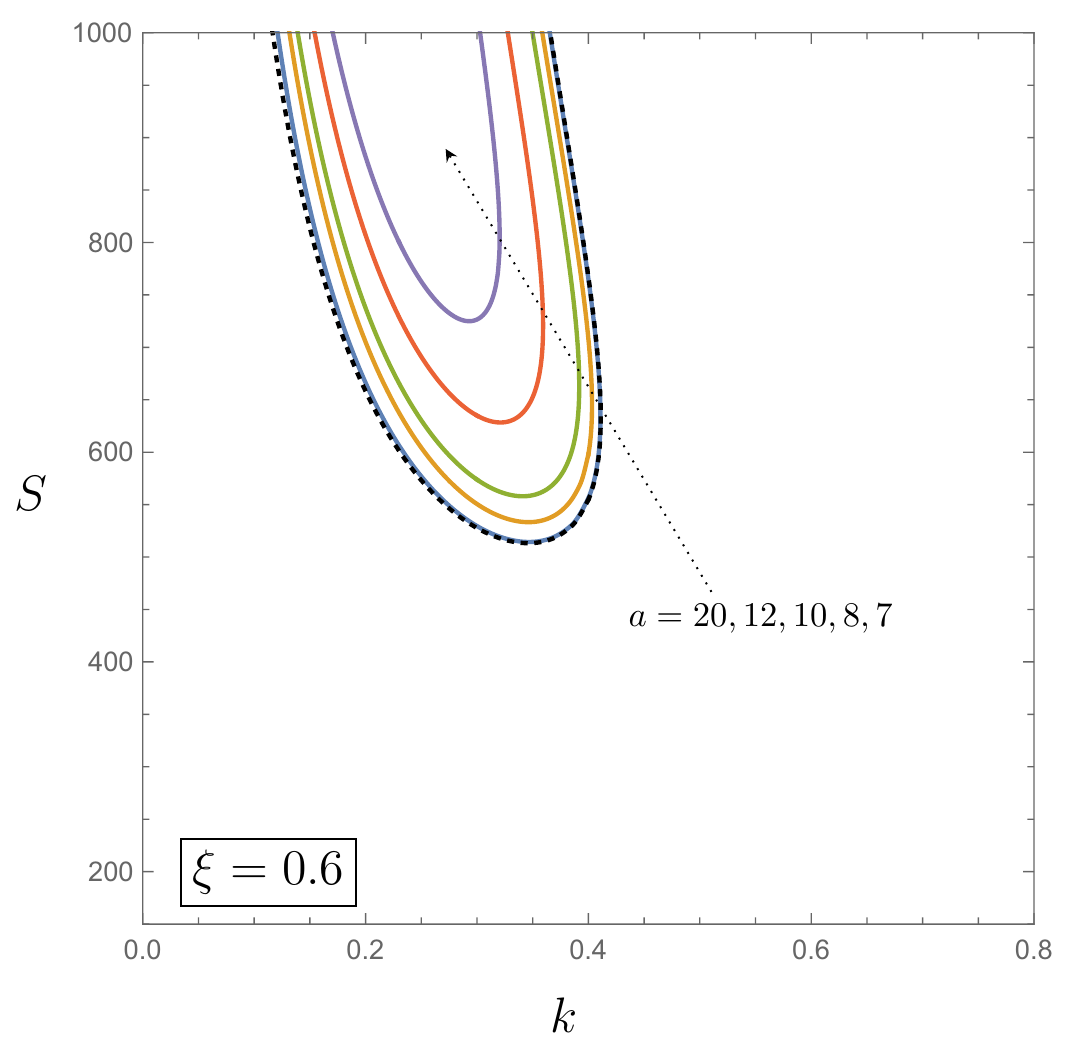}
\caption{\label{fig5}Neutral stability curves in the $(k,S)$ plane for $\xi=1$ and $\xi = 0.6$ with different values of $a$ (solid coloured curves). The dotted black curves describe the limit $a \to \infty$.}
\end{figure}

\subsection{The neutral stability curves}
The solution of \equa{35} leads to the determination of the neutral stability curves, namely the curves drawn in the $(k,S)$ plane which describe the condition of zero growth rate. In other words, the neutral stability curves are isolines of the imaginary part of the complex parameter $\omega$ corresponding to a zero value. Among the many neutral stability curves existing in the $(k,S)$ plane, our attention is focussed on that displaying the lowest values of $S$ when input values of $a$ and $\xi$ are prescribed. In fact, the lowest neutral stability curve in the $(k,S)$ plane captures the parametric condition for the initiation of the instability. As this curve usually features an absolute minimum of $S$ for a given $k$, this minimum yields the threshold for the linear convective instability. The values of $k$ and $S$ for such a minimum are called critical and denoted with $k_c$ and $S_c$.
In order to achieve these results, \equa{35} must be solved numerically. There are several techniques available for the solution of stability eigenvalue problems formulated through systems of ordinary differential equations. Comprehensive surveys and comparisons among the methods are available in \citet{dongarra1996chebyshev} and in \citet{straughan1996two}.

Our analysis of the neutral stability curves and of the critical values is carried out by employing a shooting method solution of \equa{35} for input values of $(a, \xi)$. The code employed for the implementation of such a method is just an adaption to the diverse boundary conditions of that described in \citet{Barletta2015}. In that paper, the test of the numerical accuracy for this code is also discussed. An alternative and more extensive presentation of the shooting method for linear stability eigenvalue problems is also available in \citet{barletta2019routes}.

It is desirable to see the determined critical value of $S$, for given $(a, \xi)$, formulated with the Rayleigh number $R$ in order to fully understand the mode selection at the onset of the linear instability. This information is easily gathered from \equa{25}, as this equation clearly shows that the least value of $R_c$ able to reproduce a given $S_c$ is obtained when $k_z = k$. This means that the preferred modes at the onset of instability are those with $k_y = 0$, the so-called transverse normal modes. 

The numerical solution of \equa{35} leads to a first important remark: the linear transition to instability occurs with non-travelling modes. This means that the neutral stability curves, defined by a zero imaginary part of $\omega$, are characterised also by a zero real part of $\omega$, so that the phase velocity of the neutrally stable modes is zero. This feature is justified by the numerical solution of the eigenvalue problem (\ref{35}) for several input data $(a, \xi)$.

Figures~\ref{fig3}--\ref{fig5} display different frames for different values of $\xi$. In each frame, the neutral stability curves are displayed for distinct values of $a$, with the asymptotic case $a \to \infty$ reported for comparison as a dotted black line. Figures~\ref{fig3} and \ref{fig4} also show the asymptotic case $\xi \to \infty$ as a dashed grey line for a comparison with the situation examined in \citet{Barletta2015}. Figure~\ref{fig3}, relative to $\xi = 100$ and $10$, displays an evident cluttering of the curves close to the asymptote $a \to \infty$ for the largest values of $a$. For the case $\xi=100$, the largest values of $a$ mean $a > 1.5$. In this case, there is no visible distinction between the asymptotes $\xi \to \infty$ and $a \to \infty$. We can interpret these findings by saying that, with $\xi = 100$, the analysis carried out in \citet{Barletta2015} yields a fair description of the linear onset of instability for thickness ratios $a$ down to $2$ or even smaller. Things are different when $\xi=10$ as one finds out a marked difference between the asymptotes $\xi \to \infty$ and $a \to \infty$. This phenomenon turns out to be even more evident by exploring Fig.~\ref{fig4} with the cases $\xi = 5$ and $\xi = 2$. In Fig.~\ref{fig5}, relative to $\xi = 1$ and $0.6$, the dashed grey line for $\xi \to \infty$ is not even drawn, as such cases are too far from the condition of isobaric boundary conditions, $f_1 = 0$, devised in \citet{Barletta2015}. 

There is a systematic trend gathered from Figs.~\ref{fig3}--\ref{fig5} which ought to be emphasised. The basic flow tends to be destabilised as $a$ or $\xi$ increases. This conclusion can be inferred from Figs.~\ref{fig3}--\ref{fig5} as the neutral stability curve moves upward when either the value of $a$ or $\xi$ decreases. In fact, if the neutral stability curve moves upward, larger and larger values of $S$ are needed to trigger the instability. There is a straightforward interpretation for such a trend. When $a$ gradually decreases, the overall width of the multilayer structure, $D$, tends to approach the width of the core layer, $L$. As a consequence, the stabilising effect of the impermeable external boundaries at $x=\pm D/2$, proved by \citet{gill1969proof}, tends to affect directly the interfaces at $x=\pm L/2$, so that the core layer boundaries become gradually close to impermeability and, hence, to stability for every value of $S$. Just the same condition happens when the permeability ratio, $\xi$, tends to zero as demonstrated mathematically in Section~\ref{csizero}. It is also evident from Figs.~\ref{fig3}--\ref{fig5} that the instability tends to be caused by smaller and smaller wave numbers, $k$, for decreasing values of either $a$ or $\xi$.

\begin{table}
\vspace{-3ex}
\caption{Critical values of $k$ and $S$} 
\vspace{2ex}
\centering 
\begin{tabular}{c  c c  c c  c c  c c} 
\multicolumn{1}{c}{}
& \multicolumn{2}{c}{$a\to \infty$} & \multicolumn{2}{c}{$a=20$} & \multicolumn{2}{c}{$a=10$} & \multicolumn{2}{c}{$a=5$}  \\[1ex]
\hline
$\xi$ & $k_c$ & $S_c$  & $k_c$ & $S_c$ & $k_c$ & $S_c$ & $k_c$ & $S_c$ \\[1ex] 
0.6      & 0.346043 & 513.2325 & 0.346630 & 514.3766 & 0.341008 & 557.9773 & --- & --- \\ 
0.8      & 0.428912 & 413.6608 & 0.429127 & 413.8280 & 0.429635 & 426.9214 & 0.309628 & 724.4452 \\ 
1        & 0.496592 & 357.2767 & 0.496672 & 357.3117 & 0.498337 & 362.4913 & 0.439495 & 470.0357 \\ 
2        & 0.696054 & 257.5909 & 0.696056 & 257.5912 & 0.696745 & 257.9235 & 0.689313 & 270.4812 \\ 
5        & 0.880580 & 212.7840 & 0.880580 & 212.7840 & 0.880638 & 212.8010 & 0.878152 & 214.2687 \\ 
10       & 0.961336 & 202.6737 & 0.961336 & 202.6737 & 0.961338 & 202.6766 & 0.959440 & 203.0420 \\ 
20       & 1.007717 & 199.1513 & 1.007717 & 199.1513 & 1.007714 & 199.1520 & 1.006539 & 199.2567 \\
50       & 1.038056 & 197.7046 & 1.038056 & 197.7046 & 1.038054 & 197.7047 & 1.037528 & 197.7294 \\
100      & 1.048649 & 197.3556 & 1.048649 & 197.3556 & 1.048647 & 197.3557 & 1.048375 & 197.3653 \\
$\infty$ & 1.059498 & 197.0812 & 1.059498 & 197.0812 & 1.059498 & 197.0812 & 1.059498 & 197.0812 \\ 
\hline
\end{tabular}
\label{tab1} 
\end{table}

\begin{table}
\vspace{-3ex}
\caption{Critical values of $k$ and $S$} 
\vspace{2ex}
\centering 
\begin{tabular}{c  c c  c c  c c } 
\multicolumn{1}{c}{}
& \multicolumn{2}{c}{$a=2$} & \multicolumn{2}{c}{$a=1.5$} & \multicolumn{2}{c}{$a=1.2$}  \\[1ex]
$\xi$ & $k_c$ & $S_c$  & $k_c$ & $S_c$ & $k_c$ & $S_c$ \\ [1ex] 
\hline 
5        & 0.731871 & 276.7961 & ---      & ---      & --- & --- \\
10       & 0.885192 & 218.3671 & 0.749783 & 271.2792 & --- & --- \\ 
20       & 0.962897 & 203.7273 & 0.895889 & 216.6798 & 0.663605 & 320.8838 \\
30       & 0.992229 & 200.6034 & 0.943341 & 206.8434 & 0.806683 & 246.7193 \\
40       & 1.007888 & 199.3752 & 0.968895 & 203.1681 & 0.865160 & 225.6955 \\
50       & 1.017638 & 198.7474 & 0.985154 & 201.3538 & 0.898999 & 216.2033 \\
60       & 1.024293 & 198.3750 & 0.996456 & 200.3078 & 0.921864 & 210.9843 \\
80       & 1.032791 & 197.9620 & 1.011160 & 199.1856 & 0.951664 & 205.6243 \\
100      & 1.037986 & 197.7425 & 1.020310 & 198.6121 & 0.970623 & 203.0113 \\
$\infty$ & 1.059498 & 197.0812 & 1.059498 & 197.0812 & 1.059498 & 197.0812 \\ 
\hline 
\end{tabular}
\label{tab2} 
\end{table}

\begin{figure}
\centering
\includegraphics[width=0.9\textwidth]{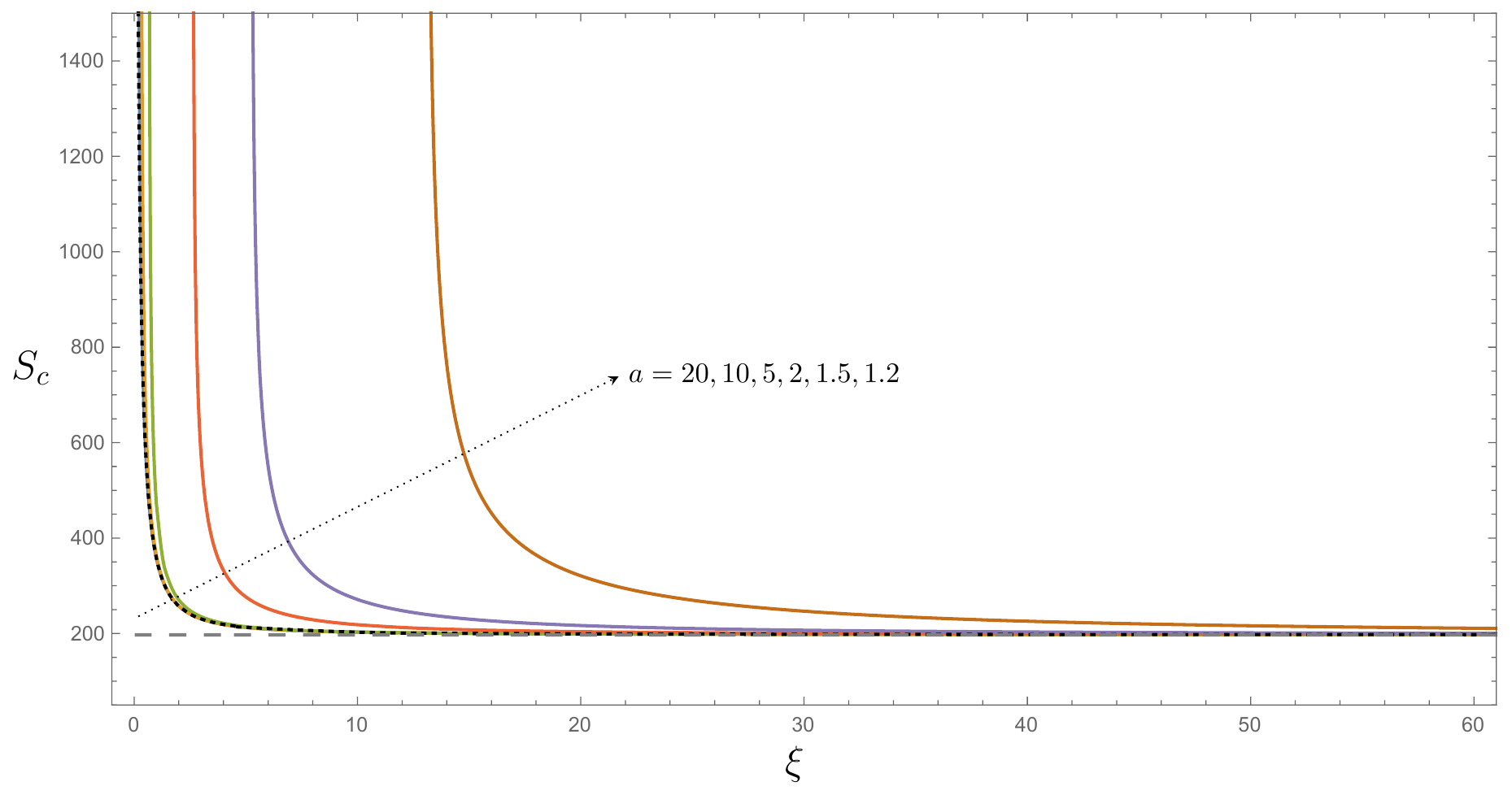}
\caption{\label{fig6}Critical value of $S$ versus $\xi$ for $a$ decreasing from $20$ to $1.2$ (solid coloured curves). The dotted black curve describes the limit $a \to \infty$, while the dashed grey line denotes the asymptotic value for $\xi \to \infty$, $S_c = 197.0812$.}
\end{figure}

\begin{figure}
\centering
\includegraphics[width=0.9\textwidth]{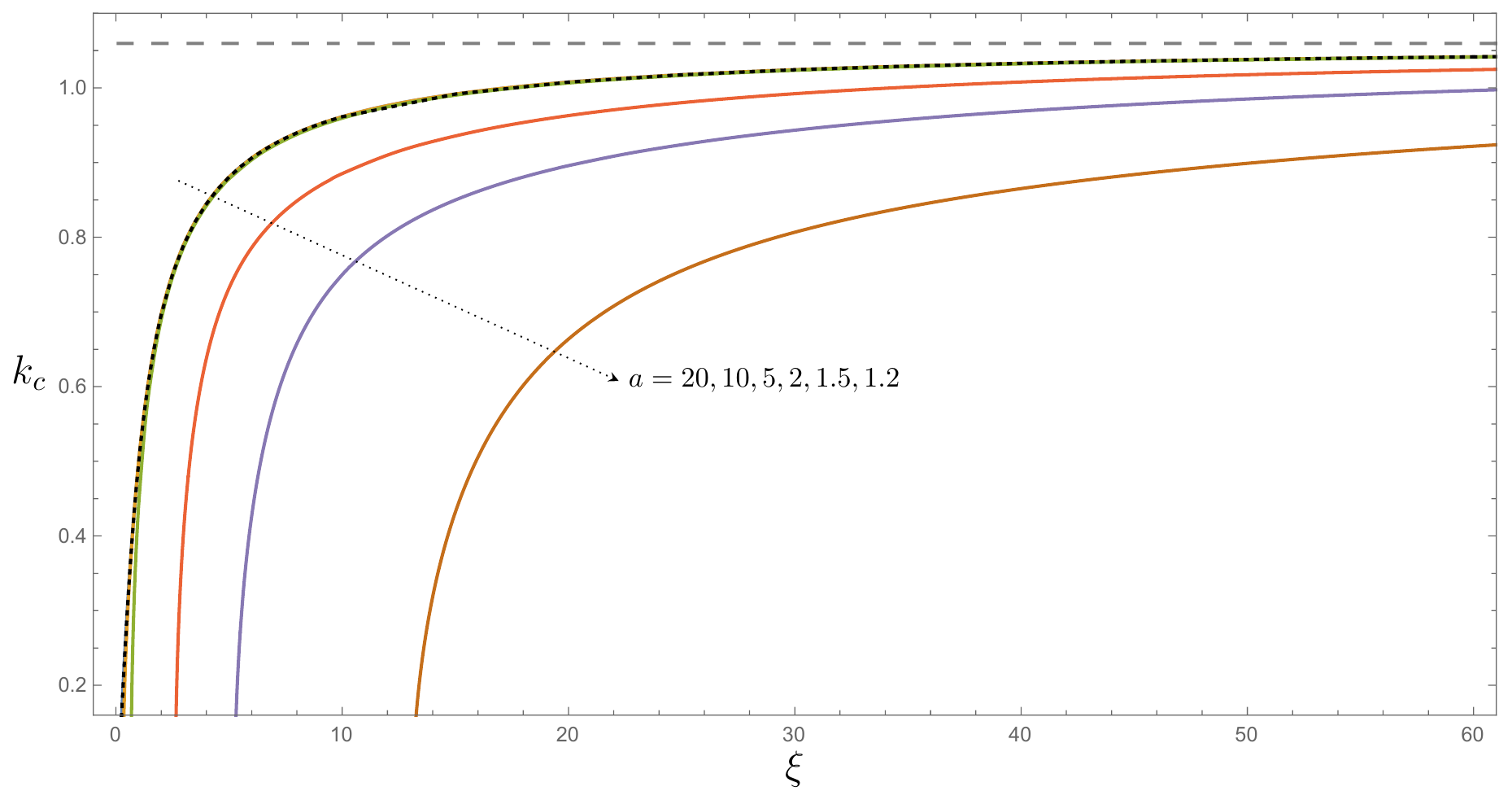}
\caption{\label{fig7}Critical value of $k$ versus $\xi$ for $a$ decreasing from $20$ to $1.2$ (solid coloured curves). The dotted black curve describes the limit $a \to \infty$, while the dashed grey line denotes the asymptotic value for $\xi \to \infty$, $k_c = 1.059498$.}
\end{figure}

\subsection{Critical conditions for the instability}
The minimum of the neutral stability curve in the $(k,S)$ plane identifies the onset of the linear instability, with the least value of $S$ needed for the convection cells to display a growing amplitude in time. Critical values $k=k_c$ and $S=S_c$ are reported in Tables~\ref{tab1} and \ref{tab2} versus $\xi$ for different decreasing values of $a$, from the asymptotic condition $a \to \infty$ to $a=1.2$. Table~\ref{tab2}, in particular, shows up the difficulty in the numerical evaluation of $(k_c, S_c)$ when $a$ is close to unity and $\xi$ is smaller than $10$. In fact, such values are for a regime where $S_c$ is significantly larger than $10^3$, so that the numerical accuracy is considerably reduced. This feature affects Table~\ref{tab1}, where the numerical data for the case $\xi=0.6$ and $a=5$ are not reported, and likewise in Table~\ref{tab2} for the cases $(\xi=5, a=1.5)$, $(\xi=5, a=1.2)$ and $(\xi=10, a=1.2)$.

The critical data reported in Tables~\ref{tab1} and \ref{tab2} are presented graphically in Figs.~\ref{fig6} and \ref{fig7} where the trends of $S_c$ and $k_c$ versus $\xi$ are displayed. These figures reveal that the cases $a=20, 10, 5$ yield hardly distinguishable curves almost overlapped to the asymptotic dotted black line corresponding to the limit $a \to \infty$. A departure from this behaviour emerges only for small values of $\xi$ in a range where the singularity of $S_c$ is displayed in Fig.~\ref{fig6} with a steep increase of the critical value of $S$ as $\xi$ decreases. We also report that the asymptotic regime $\xi \to \infty$ is approached by $S_c$ more and more rapidly the larger is the value of $a$. Thus, one can say that the case examined by \citet{Barletta2015} is one naturally emerging when both $a$ and $\xi$ are large enough. For instance, a glance at Fig.~\ref{fig6} may suggest $a \ge 5$ and $\xi \ge 20$ as a possible indication where the model of permeable boundaries employed by \citet{Barletta2015} can be considered as a fair enough approximation for practical purposes. 

\begin{figure}
\centering
\includegraphics[width=0.9\textwidth]{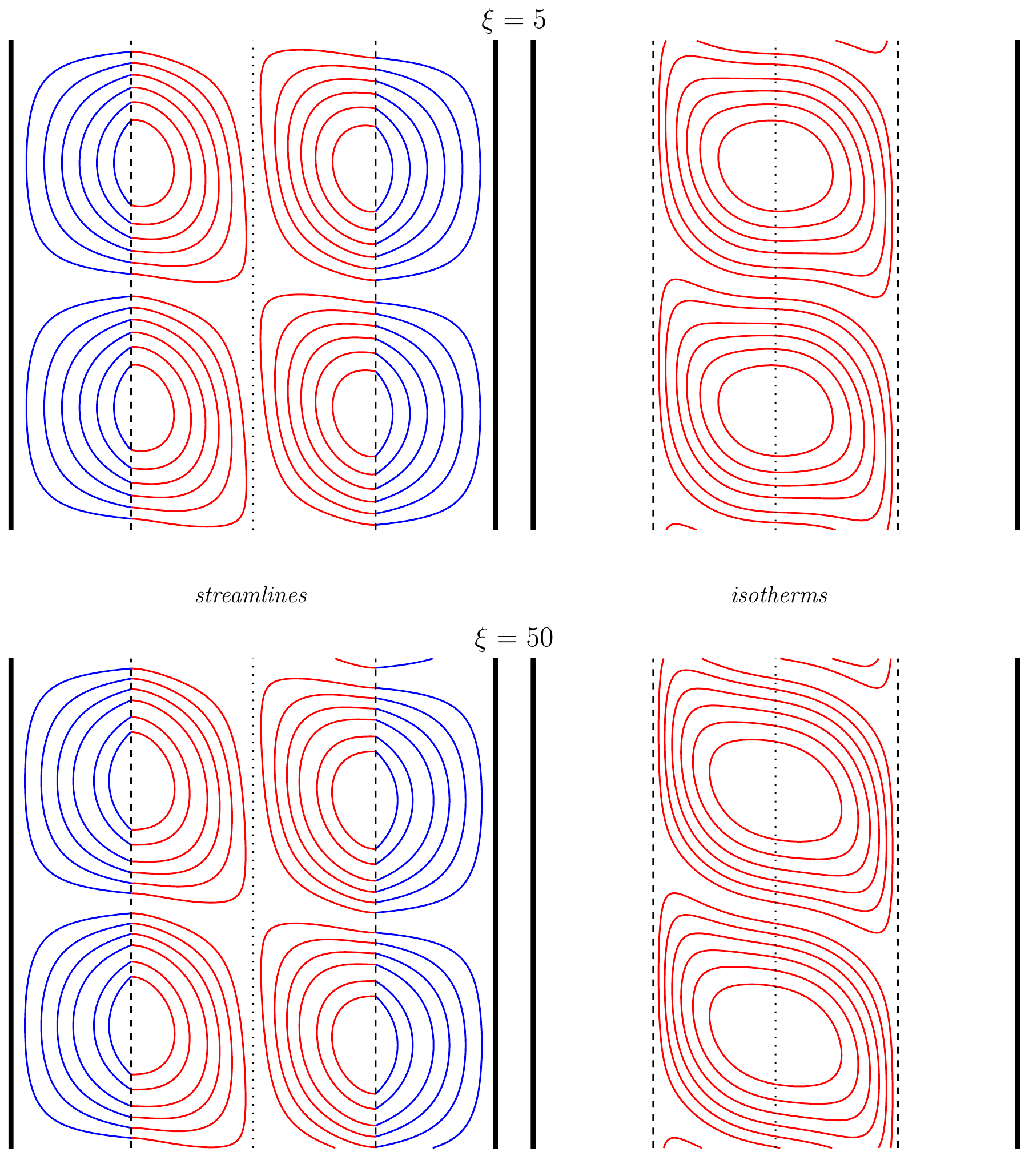}
\caption{\label{fig8}Perturbation streamlines and isotherms for transverse modes ($k_y=0$, $k_z=k$ and $S=R$) with $a=2$ and either $\xi=5$ or $\xi=50$, at critical conditions $k=k_c$ and $R=S=S_c$. The dotted black line is at $x=0$, while the dashed lines denote the interfaces $x=\pm 1/2$. The vertical range is over a period, $0 \le z \le 2\pi/k_c$.}
\end{figure}

\begin{figure}
\centering
\includegraphics[width=0.9\textwidth]{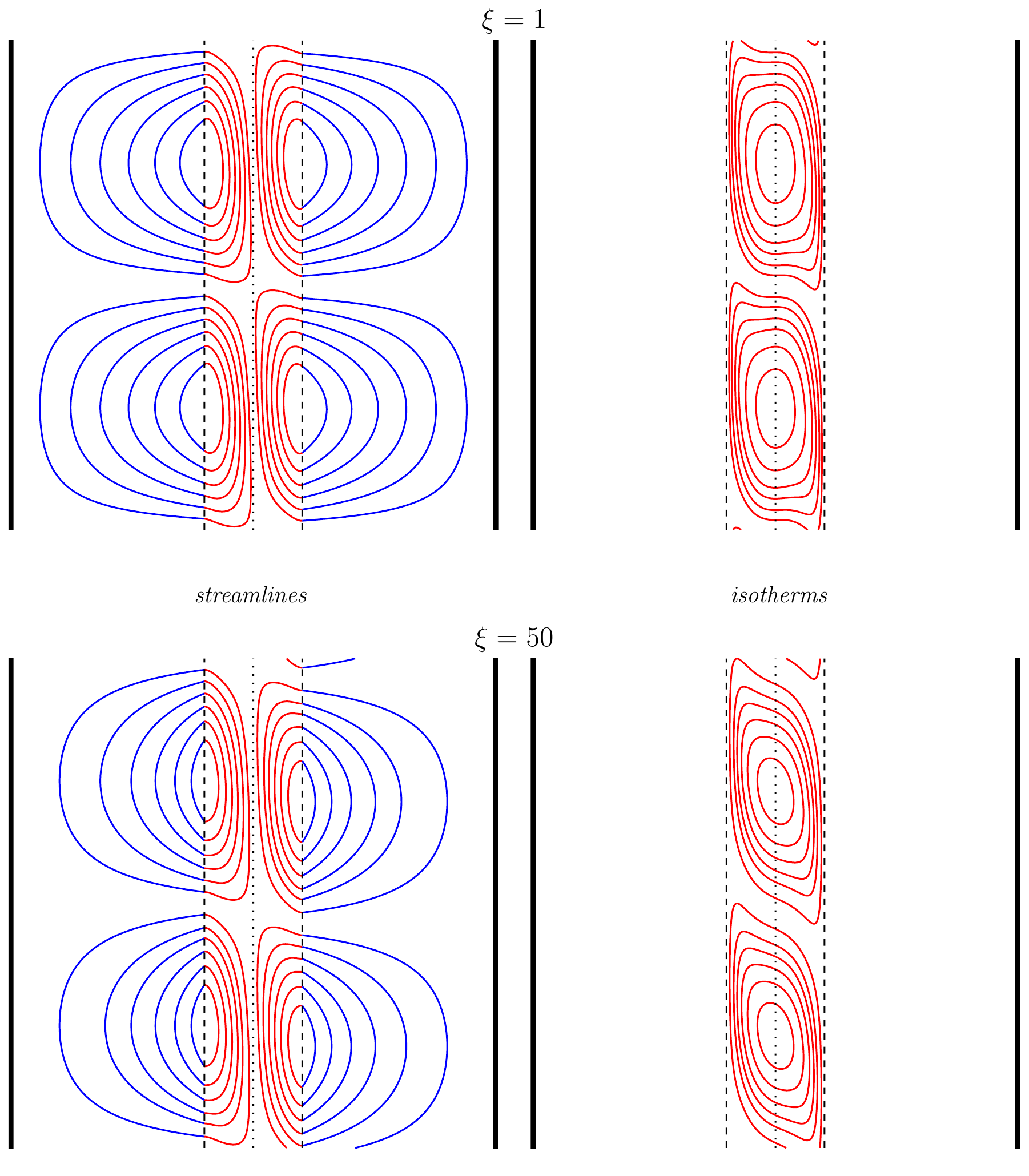}
\caption{\label{fig9}Perturbation streamlines and isotherms for transverse modes ($k_y=0$, $k_z=k$ and $S=R$) with $a=5$ and either $\xi=1$ or $\xi=50$, at critical conditions $k=k_c$ and $R=S=S_c$. The dotted black line is at $x=0$, while the dashed lines denote the interfaces $x=\pm 1/2$. The vertical range is over a period, $0 \le z \le 2\pi/k_c$.}
\end{figure}

\subsection{Streamlines and isotherms}
The visualisation of the streamlines and isotherms associated with the perturbation modes is quite important, especially at the critical conditions $k=k_c$ and $S=S_c$ defining the initiation of the convective instability. In fact, even if the flow patterns of convection are evaluated according to the linear theory, it is generally retained that such patterns are the starting point for the development of the nonlinear analysis of convection. This is particularly evident in the formulation of the weakly nonlinear theory of convective instability \citep{drazin2004hydrodynamic}. In fact, the weakly nonlinear theory studies the interaction between normal modes in the vicinity of the critical condition. 

Starting from \equasa{9}{17}, one can define the perturbation velocity,
\eqn{
\hat{\vb{u}}_1 = - \grad{\hat{P}_1} + R\, \hat{T}_1\, \vu{e}_z , 
\nonumber \\
\hat{\vb{u}}_2 = - \xi\, \grad{\hat{P}_2} + R \xi\, \hat{T}_2\, \vu{e}_z .
}{40}
We mentioned that the onset of instability occurs with transverse modes, {\em i.e.}, the two-dimensional perturbations in the $(x,z)$ plane with $k_y=0$ and $k=k_z$. Accordingly, only the velocity components $(\hat{u}_m, \hat{w}_m)$, with $m=1,2$, are nonzero for such modes.
The mass balance constrains $\hat{\vb{u}}_m$ to be solenoidal. Thus, one can introduce the streamfunction $\hat{\Psi}_m$, so that the divergence of $\hat{\vb{u}}_m$ is identically zero,
\eqn{
\hat{u}_m = \pdv{\hat{\Psi}_m}{z} \qc \hat{w}_m = -\pdv{\hat{\Psi}_m}{x}.
}{41}
Following \equa{21}, we express the streamfunction as a normal mode,
\eqn{
\hat{\Psi}_m
=
\psi_m(x) \,
e^{i \qty( k z - \omega t)}.
}{42}
Then, \equas{21} and (\ref{40})--(\ref{42}) yield
\eqn{
i k\, \psi_1(x) = - f'_1(x) \qc\hspace{2.5mm} 
- \psi'_1(x) = -i k\, f_1(x) + R\, h_1(x),
\nonumber\\
i k\, \psi_2(x) = - \xi\, f'_2(x) \qc 
- \psi'_2(x) = -i k \xi\, f_2(x) + R\xi\, h_2(x).
}{43}
As a consequence, the interface conditions at $x=\pm1/2$ expressed by \equa{24} imply a continuity of $\psi_m$ across such interfaces with a discontinuity in its first derivative when $\xi \ne 1$,
\eqn{
x = \pm 1/2 :  \qquad \psi_1 = \psi_2 , \quad \xi\, \psi'_1 = \psi'_2 .
}{44}
By employing \equasa{42}{44}, one realises that $\vb{\nabla} \hat{\Psi}_m$ undergoes a cusp discontinuity at each interface where its $x$ component changes discontinuously, if $\xi \ne 1$, while the $z$ component is always continuous.
Such cusp discontinuities of the streamlines at the interfaces are clearly visible in Figs.~\ref{fig8} and \ref{fig9}, except for the case $a=5$ and $\xi=1$ reported in the upper part of Fig.~\ref{fig9}. These figures illustrate just four sample cases in order to display how the convective instability induces a cellular flow which penetrates the ${\rm M}_2$ layers. In such external layers, the saturated porous medium is otherwise isothermal, at equilibrium with the external impermeable boundaries, $x=\pm a/2$. The latter feature emerges quite clearly from Figs.~\ref{fig8} and \ref{fig9} as the cellular patterns for the isotherms are entirely confined in the core ${\rm M}_1$ layer, that is, within the region $-1/2 \le x \le 1/2$. The red/blue colour code for the different media introduced in Fig.~\ref{fig1} is used also in Figs.~\ref{fig8} and \ref{fig9} for convenience.

\section{Conclusions}
The onset of convection in a vertical porous slab has been analysed by assuming a three-layer structure. In fact, an internal layer is sandwiched between two identical external layers having properties different from the core. The multilayer slab is saturated by a Newtonian fluid and the external layers are much more thermally conductive than the inner core. The multilayer slab is bounded by impermeable isothermal walls kept at different temperatures. 

The dynamics of convection is governed by Darcy's law and by the thermal buoyancy modelled through the Boussinesq approximation. Being extremely conductive, the external layers are thermally passive but they can be penetrated by the convection cells arising in the core layer. The basic state features a vertical flow, with a piecewise linear velocity profile, in a thermal conduction regime. Such a basic state reproduces in the inner layer that envisaged by both \citet{gill1969proof} and \citet{Barletta2015}. These authors considered a single-layer homogeneous slab with either impermeable boundaries \citep{gill1969proof} or permeable boundaries \citep{Barletta2015}.

The linear perturbations of the basic flow have been studied by employing a modal analysis. The neutral stability curves have been obtained in the $(k, S)$ plane, where $k$ is the wave number and $S$ is a suitably rescaled Rayleigh number, which coincides with the Rayleigh number $R$ in the special case of two-dimensional transverse modes. The neutral stability condition depends on two governing parameters: the permeability ratio of the outer layers to the inner layer, $\xi$; the width ratio of the whole slab to the inner layer, $a$. The critical conditions $k=k_c$ and $S=S_c$, corresponding to the point of minimum $S$ along a neutral stability curve, have been also evaluated.

The main results obtained from the linear stability analysis are the following:

\begin{itemize}
\item The most unstable perturbation modes are transverse. The transverse modes are two-dimensional lying in the vertical $(x,z)$ plane, where the $x$ axis is horizontal and perpendicular to the slab and the $z$ axis is vertical. 
\item The basic flow is destabilised both by the increase of $a$ and the increase of $\xi$. The most unstable parametric setup is the limiting case $\xi \to \infty$, where the neutral stability condition is not influenced by the value of $a$ as the stability eigenvalue problem coincides with that laid out and solved by \citet{Barletta2015}. 
\item The limiting case $\xi \to 0$ drives the basic flow to linear stability for every Rayleigh number. Such an asymptotic condition is that devised in the rigorous proof of stability presented by \citet{gill1969proof} in his classical paper. The interfaces between the internal layer and the external layers become, in fact, perfectly impermeable in this case.
\item Since the external layers have a thermal conductivity much larger than that of the core layer, they are isothermal with the same temperature as the neighbouring boundary. This happens both for the basic state and for the perturbed flow. Thus, the external layers are thermally passive at onset of instability. When the convection cells emerge at supercritical conditions, the streamlines of such cells penetrate the external layers showing up cusp discontinuities at the interfaces. The origin of the cusp discontinuities relies on the difference between the permeability of the external layers relative to that of the core layer.
\end{itemize}

The analysis carried out in this paper points out that the multilayer structure of a vertical porous slab with impermeable isothermal boundaries may induce the onset of convective instability whereas a single-layer vertical slab with the same boundary conditions displays no instability. This result points out a new connection between two apparently different types of instability: the instability in a vertical porous slab with a heterogeneous permeability structure and that in a homogeneous vertical slab with permeable boundaries. The former case, relative to a continuous type of heterogeneity, has been examined by \citet{shankar2022gill}. On the other hand, the latter case is retrieved from our analysis when the asymptotic condition $\xi \to \infty$ is examined, perfectly matching the mathematical formulation and the numerical results presented in \citet{Barletta2015}. 

\section*{Acknowledgement}
The authors acknowledge financial support from Grant No. PRIN 2017F7KZWS provided by the Italian
Ministero dell'Istruzione, dell'Universit\`a e della Ricerca.


\end{document}